\documentclass[5p,twocolumn]{elsarticle}

\usepackage{amsmath,amssymb}
\usepackage{bm}
\usepackage{hyperref}
\usepackage{setspace}
\begin{document}

\begin{frontmatter}

\title{Hyperbolic network embedding via dimensional annealing and Barnes--Hut approximation}

\author[c3s,ceu,snu]{Bukyoung Jhun}
\ead{jhun@c3s.uni-frankfurt.de}
\author[KIAS,sap]{Gangmin Son}
\author[KIAS]{Hoyun Choi}
\author[GNU]{Sang Hoon Lee}
\ead{lshlj82@gnu.ac.kr}

\address[c3s]{Center for Critical Computational Studies, Goethe-University Frankfurt, Frankfurt am Main, Germany}
\address[ceu]{Department of Network and Data Science, Central European University, Vienna, Austria}
\address[snu]{Department of Physics and Astronomy, Seoul National University, Seoul, Korea}
\address[KIAS]{School of Computational Science, Korea Institute for Advanced Study, Seoul, Korea}
\address[sap]{Dipartimento di Fisica, Sapienza Universit\`a di Roma, Rome, Italy}
\address[GNU]{Department of Physics, Gyeongsang National University, Jinju, Korea}

\begin{abstract}
Common features observed in complex networks from various domains can be explained by network geometry in hyperbolic space.
Inferring the latent coordinates of nodes enables efficient navigation, link prediction, and geometric renormalization.
However, the loss landscape of the hyperbolic model is extremely rugged due to the topological constraints of the low-dimensional embedding space. To overcome this, we introduce the dimensional annealing framework, which bypasses these constraints by introducing an auxiliary dimension.
We show that dimensional annealing achieves higher-quality embeddings than existing energy-based embedding methods.
Additionally, we incorporate the Barnes--Hut approximation for computational scalability.
\end{abstract}

\begin{keyword}
network embedding \sep hyperbolic model \sep complex networks \sep dimensional annealing \sep Barnes--Hut approximation
\end{keyword}\end{frontmatter}

\section{Introduction}

\begin{figure*}[htb]
    \centering
    \includegraphics[width=0.7\textwidth]{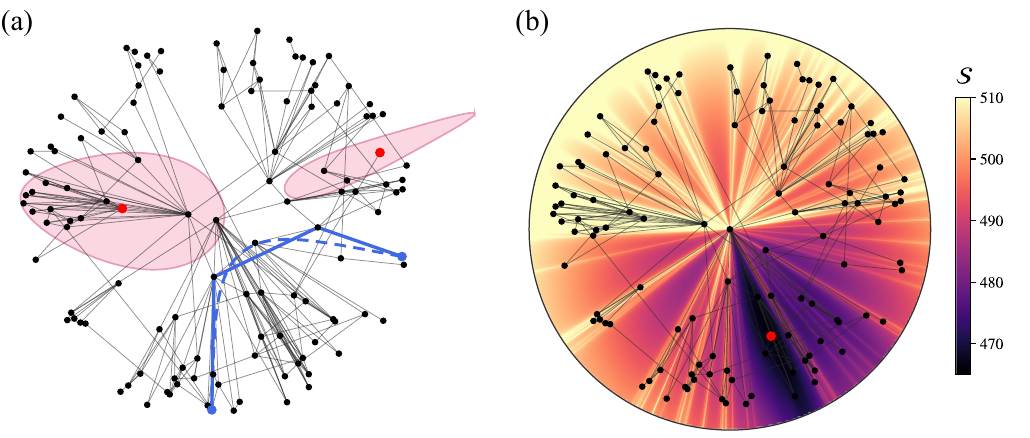}
    \caption{(a) Hyperbolic network model ($N=200$) embedded on a Poincaré disk.
    The two pink hyperbolic circles have the same radius ($R=2.5$), but contain 16 and 3 nodes, respectively.
    A circle with a fixed radius contains more nodes when it is closer to the center of the disk.
    The dashed blue curve shows the hyperbolic geodesic between the two blue dots, while the solid blue lines show the network shortest path.
    Both tend to bend towards the center of the disk, reflecting the tendency of the shortest paths to traverse high-degree nodes.
    (b) Negative log-likelihood (NLL) landscape of the same network.
    All coordinates are fixed at their original position except one node depicted as the red dot.
    There are an extensive number of local minima that are separated by potential walls stretching in the radial direction, making gradient-descent-based optimization difficult.
    }
    \label{fig:hyperbolic_model}
\end{figure*}

Many empirical networks observed in various domains exhibit heavy-tailed degree distributions, high clustering coefficients, and short diameters~\cite{albert2002statistical}.
One prominent explanation for these observations is that the network structure arises from a latent geometry, in which the connection probability between two nodes depends on their distance in a low-dimensional metric space~\cite{barthelemy2011spatial,dall2002random,masuda2005geographical}.
If each node has a latent coordinate, and the connection probability is high between nodes that are close in the latent space, the high clustering coefficient naturally arises.
Heterogeneous degree distributions and short diameters can be explained when the latent space is hyperbolic.
In the Poincaré disk, nodes closer to the center have more neighbors, while most nodes in the periphery have a small number of neighbors (Fig.~\ref{fig:hyperbolic_model}(a)).
In this framework, a hyperbolic network model with a Fermi--Dirac distribution of connection probability has been introduced to explain both the power-law degree distribution and high clustering coefficient~\cite{boguna2021network,papadopoulos2014network,krioukov2016clustering,serrano2008self,krioukov2010hyperbolic}.
Recent studies have further demonstrated latent geometry as the origin of nestedness in real-world hypergraphs~\cite{nortier2026latent}.

In real-world networks, the latent hyperbolic coordinates are not available and must be inferred from the network structure alone.
The importance of such an inverse problem is that the hyperbolic model is not merely a visualization tool.
When hyperbolic coordinates are successfully inferred, one can navigate the network without global information~\cite{krioukov2010hyperbolic,boguna2010sustaining}.
Moreover, hyperbolic coordinates can be used for community detection~\cite{kovacs2021inherent}, link prediction~\cite{wang2016link,kitsak2020link}, network dismantling~\cite{osat2023embedding}, and geometric renormalization group~\cite{garcia2018multiscale,kolk2024renormalization}.

Several algorithms for this inverse problem, or \emph{hyperbolic embedding}, have been proposed.
The quality of such embeddings can be evaluated by the negative log-likelihood (NLL) of the observed network.
One fundamental challenge in hyperbolic embedding is that the NLL landscape is extremely rugged, with a large number of local minima separated by infinite potential walls.
This ruggedness stems from the fact that when the hyperbolic coordinates of two unconnected nodes approach each other, the NLL diverges to infinity.
Since the similarity space of the hyperbolic network model is one-dimensional, two points cannot change their order without their trajectories crossing each other.
This leads force-based embedding methods, which rely on gradient descent to minimize the NLL, to be trapped in local minima.
Instead, current literature employs energy-based optimization schemes, in which the hyperbolic coordinates are perturbed by a finite amount, and the proposed perturbations are accepted or rejected based on the resulting change in NLL.
Recent Bayesian approaches have revealed strong multimodality and potential barriers in the posterior landscape~\cite{lizotte2025symmetry}.

Here, we introduce the \emph{dimensional annealing} framework, which, instead of altering the optimization algorithm or the objective function of the problem, extends the domain of the embedding.
By introducing an auxiliary dimension, two nodes can exchange their coordinates without colliding, thereby avoiding the divergence of the NLL.
To recover an embedding in the original low-dimensional space, we gradually confine the auxiliary coordinate to zero.

We implement this framework in \textbf{D}imensional \textbf{A}nnealing--based \textbf{Hy}perbolic \textbf{E}mbedding (DAHyE), a C++ library with a Python wrapper for obtaining high-quality hyperbolic embeddings of networks.
It also utilizes the Barnes--Hut approximation that groups the nodes with small and smooth contributions into a single cluster, reducing the computational cost to $\mathcal{O}(N\log N)$.

By testing this framework on synthetic and real-world networks from diverse domains, we show that dimensional annealing produces higher-quality embeddings than the state-of-the-art Mercator embedding~\cite{garcia2019mercator,jankowski2023d} in most cases.
Additionally, we show that the Barnes--Hut approximation drastically improves the scalability of the method by decreasing the computational complexity to $\mathcal{O}(N \log N)$.

This article is organized as follows: In Sec.~\ref{sec:background}, we introduce the hyperbolic network model and explain the inverse problem, hyperbolic greedy navigation~\cite{boguna2009navigating}, and the A* search algorithm. In Sec.~\ref{sec:dimensional_annealing}, the dimensional annealing framework is introduced. The Barnes--Hut approximation is explained in Sec.~\ref{sec:barnes_hut}. The quality of the embedding and the scalability are assessed in Sec.~\ref{sec:results}. The conclusions are provided in Sec.~\ref{sec:conclusion}.

\section{Hyperbolic network model}
\label{sec:background}

In the $\mathbb{S}^1$ model, each of the $N$ nodes in the network has a coordinate $\theta_i \in [0,2\pi)$ on a circle (similarity space) of radius $\mathcal{R}=N/(2\pi)$.
Each node also has a hidden degree $\kappa_i > 0$ associated with its tendency to have a high degree.
The probability $p_{ij}$ that two nodes $i$ and $j$ are connected is related to the distance between the two nodes $d_{ij} = \mathcal{R}\Delta \theta_{ij}$ on the similarity space.
\begin{align}
p_{ij} &= \frac{1}{1 + \chi_{ij}^\beta} \label{eq:p_ij} \,, \\
\chi_{ij} &= \frac{N}{2\pi} \frac{\Delta \theta_{ij}}{\mu \kappa_i \kappa_j} \,. \label{eq:chi_ij}
\end{align}
Here, $\Delta \theta_{ij}=\pi-|\pi-|\theta_i-\theta_j|| \in [0,\pi]$ is the angular distance between the two nodes. $\beta$ is the inverse temperature that controls the clustering coefficient, and $\mu$ is the parameter that controls the average degree of the generated network.
We set $\mu=\frac{\beta}{2\pi\langle k\rangle} \sin (\pi/\beta)$ so that the expected degree of each node matches the hidden degree parameter: $k_i \simeq \kappa_i$ ($\beta>1$).

\subsection{Hyperbolic embedding}
\label{sec:hyperbolic_embedding}

For real networks, we only have access to the resulting network structure, not its hyperbolic coordinates $(\theta_i, \kappa_i)$.
Therefore, we need to infer the hyperbolic coordinates from the network structure.
Let $\bm{\Theta} = (\{\theta_i\}, \{\kappa_i\}, \beta, \mu)$ be the parameters of the hyperbolic model.
A standard approach to inferring $\bm{\Theta}$ from a given network is maximum a posteriori estimation.
According to Bayes' theorem, 
\begin{align}
P(\bm{\Theta} | A) = \frac{P(A | \bm{\Theta}) P(\bm{\Theta})}{P(A)} \propto P(A | \bm{\Theta}) P(\bm{\Theta}) \,.
\end{align}
where $A$ is the adjacency matrix.
With a prior $P(\bm{\Theta})$ that is constant over the admissible parameter domain, maximum a posteriori estimation is equivalent to maximum likelihood estimation
\begin{align}
\hat{\bm{\Theta}} = \arg\max_{\bm{\Theta}} P(\bm{\Theta} | A) = \arg \max_{\bm{\Theta}} P(A | \bm{\Theta}) \,.
\end{align}

The likelihood of the observed network is computed as
\begin{align}
\mathcal{L} = P(A | \bm{\Theta}) = \prod_{i<j} p_{ij}^{A_{ij}} (1-p_{ij})^{1-A_{ij}} \,. \label{eq:likelihood} 
\end{align}
Maximizing this likelihood is equivalent to minimizing the NLL
\begin{align}
\mathcal{S} = -\ln \mathcal{L} = -\sum_{(i,j)\in E} \ln p_{ij} - \sum_{(i,j)\notin E} \ln (1-p_{ij}) \,, \label{eq:negative_log_likelihood}
\end{align}
where $E$ is the set of edges in the network.
In this process, the first term acts as an attractive force that draws connected nodes closer, while the second sum is a repulsive term that pushes unconnected nodes apart.
Hyperbolic embedding methods seek to minimize the NLL in Eq.~\eqref{eq:negative_log_likelihood}.

Note that the NLL is invariant under rotation $\theta_i \mapsto \theta_i + \phi$ and reflection $\theta_i \mapsto -\theta_i$.
As a consequence, the optimal embedding is not unique but belongs to an equivalence class of symmetry-related embeddings with identical NLL.
In practice, embedding methods break this symmetry and find only one with a broken symmetry.

\subsection{Laplacian eigenmap initialization}
The NLL \eqref{eq:negative_log_likelihood} is not convex with respect to $(\theta_i, \kappa_i)$, and local optima are continuously connected due to the symmetry.
To initialize the embedding coordinates close to their optimal values at vastly lower computational cost, we employ Laplacian eigenmap~\cite{alanis2016efficient,muscoloni2017machine}.

Laplacian eigenmap embeds high-dimensional points $\bm{x}_i \in \mathbb{R}^n$ onto a lower-dimensional space $\bm{y}_i \in \mathbb{R}^m$ while preserving local proximity, so that the neighboring points in the original space remain close.
Hence, it minimizes the loss function
\begin{align}
\mathcal{E} = \sum_{i,j} || \bm{y}_i - \bm{y}_j||^2 \omega_{ij} \label{eq:loss_function} \,,
\end{align}
where $\omega_{ij}$ denotes the weight factor representing how close the nodes $i$ and $j$ are in the original space.
Generally, a heat kernel of the form
\begin{align}
\omega_{ij} = \exp \left( - \frac{|| \bm{x}_i - \bm{x}_j ||^2}{t} \right)
\end{align}
is used.
If two nodes are close in the original space, $\omega$ is large (close to 1); therefore, placing them far apart in the lower dimension rapidly increases the loss function.
As the loss function $\mathcal{E} \geq 0$ has a trivial solution $\bm{y}_1 =  \bm{y}_2 = \cdots = \bm{y}_N$, we employ the normalization condition
\begin{align}
Y^T D Y = I \label{eq:laplacina_constraint1} \,,
\end{align}
where $D_{ii} =  \sum_j \omega_{ij}$.
Additionally,  we assume
\begin{align}
\sum_{ij} y_i \omega_{ij} = 0 \label{eq:laplacina_constraint2}
\end{align}
to fix the center of mass at the origin.
The minimization of the loss function Eq.~\eqref{eq:loss_function} under constraints Eq.~\eqref{eq:laplacina_constraint1} and \eqref{eq:laplacina_constraint2} is equivalent to a generalized eigenvalue problem
\begin{align}
L \bm{v} = \lambda D \bm{v} \,,
\end{align}
where
\begin{align}
    L_{ij}=
\begin{cases}
\displaystyle\sum_k\omega_{ik}, & i=j,\\
-\omega_{ij}, & i\ne j.
\end{cases}
\end{align}

This eigenmap cannot be directly used to reduce the dimension of a given network, since, unlike the usual problem of dimensional reduction, $||\bm{x}_i - \bm{x}_j||$ is not given.
We use the expected angular separation between two nodes instead of this value.
\begin{align}
\langle \Delta\theta_{ij} \rangle = \mathbb{E}[\Delta\theta_{ij} | A_{ij} = 1, \kappa_i, \kappa_j] \,.
\end{align}
We transform this value to the chord length on a circle to obtain $\omega_{ij}$
\begin{align}
d_{ij} &= 2 \sin \left( \frac{\langle \Delta\theta_{ij} \rangle}{2} \right) \,, \\
\omega_{ij} &= A_{ij} \exp\left(-\frac{d_{ij}^2}{t} \right) \,.
\end{align}
We obtain eigenvectors corresponding to the two smallest non-zero eigenvalues, $\bm{v}_1$ and $\bm{v}_2$, to construct the Laplacian eigenmap approximation of the hyperbolic angular coordinates
\begin{align}
\theta_i^{\rm LE} = {\rm atan2} (v_{2,i}, v_{1,i}) \,,
\end{align}
where ${\rm atan2} (y, x)$ is the two-argument arctangent function, which returns the polar angle of the vector $(x,y)$.

\subsection{Inverse temperature}

The inverse temperature $\beta$ also needs to be inferred from the network structure.
Here, we use the method used in the Mercator embedding~\cite{garcia2019mercator,jankowski2023d}.
For two nodes $i$ and $j$ with latent degree $\kappa_i$ and $\kappa_j$, averaging over uniformly distributed angular coordinates gives the connection probability
\begin{align}
\bar{p}(\kappa_i, \kappa_j) &= \frac{1}{\pi} \int_{0}^{\pi} \frac{d\Delta\theta}{1 + (N\Delta\theta/(2\pi\mu\kappa_i\kappa_j))^\beta} \\
&= {}_2F_1 \left(1, \frac{1}{\beta}; 1+\frac1\beta; -\Bigl(\frac{N}{2\mu\kappa_i\kappa_j}\Bigr)^\beta \right) \,,
\end{align}
where ${}_2F_1$ is the hypergeometric function
\begin{align}
    {}_2F_1(a,b;c;z) = \sum_{n=0}^{\infty} \frac{(a)_n (b)_n}{(c)_n} \frac{z^n}{n!}
\end{align}
with the (rising) Pochhammer symbol $(q)_n = q(q+1)\cdots(q+n-1)$ for $n>0$ and $(q)_0 = 1$.

Therefore, the inverse temperature $\beta$ controls the sensitivity of the connection probability to distance, affecting the clustering coefficient of the network.
When $\beta$ is large, only pairs of nodes that are near are connected, forming a large number of triangles and a high clustering coefficient.
Hence, the clustering coefficient of the network provides information for inferring $\beta$.
The model clustering coefficient $c_{\rm model}(\beta)$ can be calculated by selecting a node and its two neighbors, then computing the connection probability between the two.
The cumulative distribution of angular distance $\Delta\theta$ between two connected nodes is
\begin{align}
F(\theta) &= \frac{\int_0^\theta p(\Delta\theta) d\Delta\theta}{\int_0^\pi p(\Delta\theta) d\Delta\theta} \\
&= \frac{\theta}{\pi} \frac{{}_2F_1(1, 1/\beta; 1+1/\beta; -(N\theta/(2\pi\mu\kappa_i\kappa_j))^\beta)}{{}_2F_1(1, 1/\beta; 1+1/\beta; -(N/(2\mu\kappa_i\kappa_j))^\beta)} \,.
\end{align}
Because $F(\theta)$ monotonically increases from $F(0)=0$ to $F(\pi)=1$, we can sample $\Delta\theta$ of the two neighbors efficiently.
The model clustering coefficient can be computed by sampling $\theta(u)$ and $\theta(v)$ and summing up the contributions of the probability of triangle formation $c_k = \langle p_{uv} \rangle$.
\begin{align}
c_{\rm model}(\beta) = \frac{1}{N} \sum_{k} N_k c_k \,,
\end{align}
where $N_k$ is the number of nodes with degree $k$.
Since $c_{\rm model}(\beta)$ is a monotonically increasing function of $\beta$, we determine $\beta$ by binary search to match the clustering coefficient of the observed network.

\subsection{Greedy navigation}

One of the most prominent applications of the hyperbolic embedding is navigation.
Conventional routing to a given destination generally requires global information about the network structure.
In contrast, greedy navigation exploits node coordinates and requires only local information~\cite{kleinberg2000navigation,lee2012exploring,sun2010greedy}.
Specifically, hyperbolic greedy navigation towards destination $t$ proceeds at each step to the neighboring node $i$ that minimizes
\begin{align}
    \chi_{it} = \frac{\Delta \theta_{it}}{\mu \kappa_i \kappa_t} \,.
\end{align}
Since the angular distance is discounted by the latent popularity of the nodes, the trajectory prefers passing through hub nodes.
This matches the pattern of shortest paths in complex networks (Fig.~\ref{fig:hyperbolic_model}(a)).
Hence, neither routing tables nor global searches of the network are required, and the amount of information that must be stored in each node becomes independent of the network size.
For this reason, hyperbolic greedy navigation has attracted attention as a scalable routing approach for the Internet~\cite {boguna2010sustaining}.

Greedy navigation is effective only when the inferred coordinates faithfully encode the latent geometries, such that approaching the destination in the embedding space also tends to reduce the network distance to it.
However, this simple navigation sometimes becomes trapped when the next hop would revisit an already visited node.
We therefore use the \emph{greedy navigation success rate} $s$, measured over randomly selected source-destination pairs as a practical metric of hyperbolic embedding quality.

\subsection{A* algorithm}

Failure to find a path is often intolerable in real systems.
Therefore, the A* search algorithm~\cite{hart1968formal} has been proposed, which guarantees to find a path between two nodes whenever one exists.
Suppose we are navigating from node $s$ to node $t$.
A* evaluates each candidate node $i$ by
\begin{align}
f(i) = d(s, i) + \chi_{it} \,,
\end{align}
where $d(s,i)$ is the shortest path distance from the source $s$ found so far, and $\chi_{it}$ serves as the estimate of the remaining distance to the destination $t$.
Nodes with smaller $f$ are expanded first, so the search is guided towards the destination rather than spreading uniformly as in breadth-first search (BFS).

As in greedy navigation, a more accurate distance estimate $\chi_{it}$ can improve the search efficiency of A*.
We quantify the search cost relative to BFS using
\begin{align}
R = \frac{\langle n_{\rm A*}\rangle}{\langle n_{\rm BFS}\rangle} \,,
\end{align}
where $\langle n_{\rm A*}\rangle$ and $\langle n_{\rm BFS}\rangle$ denote the number of nodes explored by the respective algorithms, and the averages are taken over source-destination pairs.

\section{Dimensional annealing}
\label{sec:dimensional_annealing}

\begin{figure*}[tb]
    \centering
    \includegraphics[width=0.9\textwidth]{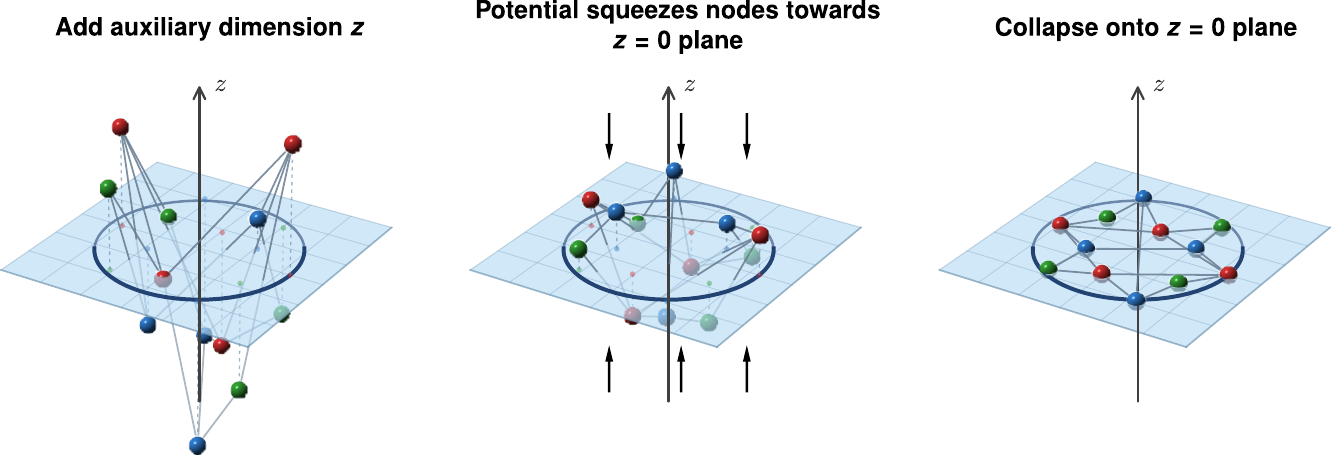}
    \caption{
    Illustration of dimensional annealing.
    First, an auxiliary dimension is introduced.
    Unlike in the original hyperbolic model, two unconnected nodes can bypass each other in the $z$-direction without a collision.
    Then, the $z$-directional potential squeezes nodes towards the $z=0$ plane.
    The nodes approach their approximate embedding coordinates utilizing the auxiliary dimension.
    Finally, the nodes collapse onto the $z=0$ plane to obtain a hyperbolic embedding.}
    \label{fig:dimensional_annealing}
\end{figure*}

The NLL $\mathcal{S}$ in Eq.~\eqref{eq:negative_log_likelihood} is differentiable with respect to coordinates and can, in principle, be optimized via gradient descent.
However, its landscape is extremely rugged with numerous local optima separated by infinite potential walls (Fig.~\ref{fig:hyperbolic_model}(b)).
For unconnected nodes $i$ and $j$, $p_{ij}$ approaches unity as their angular separation vanishes, causing the repulsive contribution $-\ln (1-p_{ij})$ to diverge.
These barriers prevent gradient descent from escaping local minima, limiting its ability to find a high-quality embedding. 

This limitation arises from the one-dimensional similarity space, where nodes cannot exchange their order without colliding.
To this end, we propose the \emph{dimensional annealing} framework to bypass this challenge by introducing an auxiliary dimension $z$ in the similarity space.
The embedding coordinates of each node are extended to $(\kappa_i, \theta_i, z_i)$.
Then, Eq.~\eqref{eq:chi_ij} is modified to
\begin{align}
    \chi_{ij}^\prime &= \frac{1}{2\pi} \frac{d_{ij}^\prime}{\mu\kappa_i\kappa_j} \,, \\
    d_{ij}^\prime &= \sqrt{(N\Delta\theta)^2 + (z_i-z_j)^2} \,.
    \label{eq:d_ij_again}
\end{align}
The NLL remains differentiable, and we minimize it using gradient descent with an additional $z$-directional potential $V(\mathbf{z},t) = a(t)\sum_i |z_i|$.
Here, the annealing coefficient $a(t) = t/T_{\rm penalty} = \ell\delta t/T_{\rm penalty}$, where $\ell$ is the number of optimization steps and $\delta t$ is the learning rate of the gradient descent.
Early in the optimization, a small $a(t)$ lets the nodes exploit the auxiliary dimension to exchange their relative order without collisions, while a growing $a(t)$ squeezes all nodes towards the $z=0$ plane so that the final embedding collapses onto $(\theta, z=0)$.
An overview of dimensional annealing is illustrated in Fig.~\ref{fig:dimensional_annealing}.

\section{Barnes--Hut approximation}
\label{sec:barnes_hut}

\begin{figure}[tb]
    \centering
    \includegraphics[width=0.85\columnwidth]{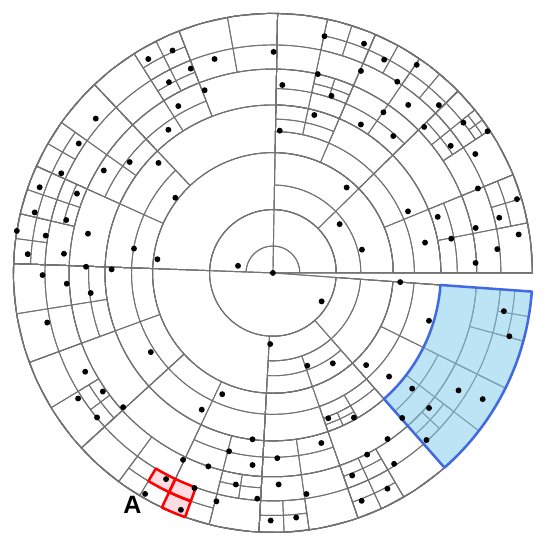}
    \caption{Illustration of Barnes--Hut approximation.
    The hyperbolic space is iteratively partitioned: when more than one node is in a cell, the cell is divided into four until each cell contains at most one node.
    When computing the negative log-likelihood (NLL) contributions of nodes that are close, the coordinates of each node are taken into account accurately.
    However, contributions of nodes that are far away are collapsed into a group, and only their center of mass is considered.
    This reduces the computational complexity of NLL computation from $\mathcal{O}(N^2)$ to $\mathcal{O}(N\log N)$.
    For instance, when computing the NLL term for node A, nodes in red cells are close; hence, each of the three cells is considered accurately in the computation.
    Meanwhile, the nodes in the blue cell are far away, and they are considered only as a group.}
    \label{fig:barnes_hut}
\end{figure}

\emph{Dimensional annealing} mitigates the problem of potential walls but leaves the $\mathcal{O}(N^2)$ complexity of the NLL because Eq.~\eqref{eq:negative_log_likelihood} contains summation over all $\binom{N}{2}$ node pairs.
As in gravitational $N$-body simulation in astronomy, we employ the \emph{Barnes--Hut approximation}~\cite{barnes1986hierarchical, hamada2009novel} to group distant nodes and approximate their combined contributions, reducing the computational complexity to $\mathcal{O}(N\log N)$.

Equation~\eqref{eq:negative_log_likelihood} can be rewritten as
\begin{align}
    \mathcal{S}
    =
    - \underbrace{\sum_{(i,j)\in E}
    [\ln p_{ij}-\ln(1-p_{ij})]}_{\text{attractive, edges only}}
    -
    \underbrace{\sum_{i<j}\ln(1-p_{ij})}_{\text{repulsive, all pairs}} \label{eq:negative_log_likelihood_2} \,.
\end{align}
The first term of Eq.~\eqref{eq:negative_log_likelihood_2} can be computed at low cost owing to network sparsity, while the second is approximated using the Barnes--Hut method.

In dimensional annealing, each node has coordinates $\theta_i$, $\kappa_i$, and $z_i$.
The coordinate space is partitioned in the $(\theta, \kappa)$ plane.
The auxiliary coordinate $z$ approaches zero as the dynamics progresses.
We do not partition the space along the $z$ direction.
Instead, only the average $z$ value in each cell is tracked.

The partition is done hierarchically using a tree structure.
The root cell encompasses all nodes.
If two or more nodes are in a cell, the cell is subdivided into four by bisecting both the $\theta$ and $\kappa$ dimensions.
The division is done iteratively until each leaf cell contains at most one node.
The Barnes-Hut tree is illustrated in Fig.~\ref{fig:barnes_hut}.
Each cell (leaf or non-leaf) keeps track of the number of nodes $n_c$ and its center of mass
\begin{align}
(\bar{\theta}_c, \bar{\kappa}_c, \bar{z}_c) = \frac{1}{n_c} \sum_{j \in c} (\theta_j, \kappa_j, z_j) \,.
\end{align}
The tree is constructed at each time step of the optimization.

Once the space is partitioned, the NLL and its gradient can be computed by traversing the tree starting from its root.
When cell $c$ is considered, we either approximate its contribution using the cell's center of mass or recursively examine its child cells.
Suppose the $\theta$-directional length of cell $c$ is $s_c$.
With precision parameter $\Theta$, if
\begin{align}
s_c/\Delta\theta_{ic} \leq \Theta \label{eq:barnes_hut_condition} \,, 
\end{align}
then we approximate that the entire contribution of the cell comes from the center of mass of cell $c$.
Otherwise, we open the cell to consider its child cells.
A large $\Theta$ increases the computational speed but decreases the accuracy, and vice versa.

If nodes are spread across the embedding space, the cells satisfying Eq.~\eqref{eq:barnes_hut_condition} are hierarchically arranged according to the distance, and the cells considered for computation of NLL remain constant at each level.
Since the tree depth scales as $\mathcal{O}(\log N)$, the total computational cost for computing the repulsive terms becomes $\mathcal{O}(N \log N)$.
Additionally, the computation of attractive terms requires $\mathcal{O}(|E|) = \mathcal{O}(N\langle k \rangle)$, which results in the final computational complexity of $\mathcal{O}(N \log N)$ for sparse networks.

\section{Results}
\label{sec:results}

\subsection{Quality of embedding}
\begin{figure}[tb]
    \centering
    \includegraphics[width=0.85\columnwidth]{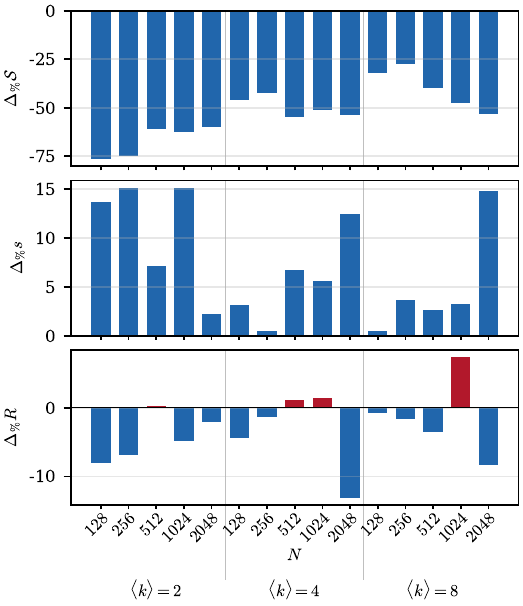}
    \caption{Comparison between the dimensional annealing framework and the state-of-the-art Mercator embedding.
    Average degree $\langle k\rangle = 2,4,8$ and $N=128, 256, \cdots, 2048$ are considered.
    Top panel illustrates the negative log-likelihood (NLL), middle panel shows greedy navigation success rate, and bottom panel shows the expansion ratio of A* search algorithm.
    Their percentage differences are plotted.
    Cases where the proposed method achieves better results are depicted in blue, and cases where the Mercator shows better results are illustrated in red.
    The proposed method exhibits better results in NLL and greedy navigation success rate for all cases.
    For most networks, the proposed method performs better in terms of A* expansion ratio.}
    \label{fig:synthetic_networks}
\end{figure}

\begin{figure*}[tb]
    \centering
    \includegraphics[width=0.9\textwidth]{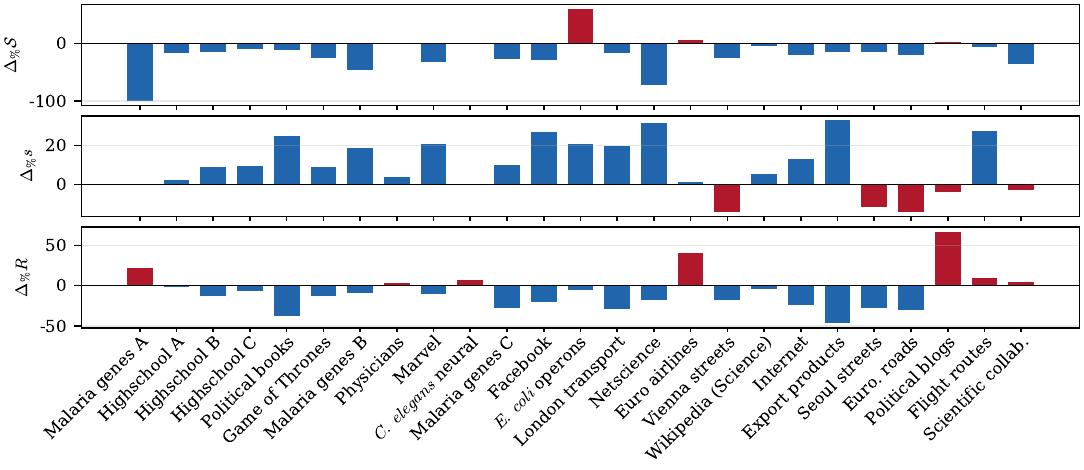}
    \caption{Comparison of dimensional annealing embedding and Mercator in real networks.
    Networks are ordered by increasing network size $N$.
    Top panel illustrates the negative log-likelihood (NLL), middle panel shows the greedy navigation success rate, and bottom panel depicts the expansion ratio of the A* search algorithm.
    Percent differences are plotted.
    Blue denotes the cases where dimensional annealing performs better than the state-of-the-art Mercator embedding, and red denotes the cases where Mercator performs better.
    In all three metrics, dimensional annealing shows better results in most cases.}
    \label{fig:real_networks}
\end{figure*}

In this section, we evaluate the effectiveness and accuracy of the dimensional annealing approach in synthetic hyperbolic models and real-world networks.
The NLL $\mathcal{S}$ in Eq.~\eqref{eq:negative_log_likelihood} directly measures how well an embedding explains the given network structure.
The greedy navigation success rate $s$ and the relative search cost $R$ of the A* search algorithm assess its practical utility for navigation.
The state-of-the-art Mercator embedding~\cite{garcia2019mercator} is used as a benchmark.

We generated synthetic hyperbolic networks with sizes $N=128, 256, \dots, 2048$, mean degrees $\langle k \rangle = 2, 4, 8$ with $\beta=4.0$.
Because Mercator embedding is designed for connected networks, we take the largest connected component of each network.
We compare the dimensional annealing and Mercator using the percentage difference
\begin{align}
\Delta_{\%} \mathcal{X} = 100 \times \frac{\mathcal{X}_{\rm DA} - \mathcal{X}_{\rm mercator}}{\mathcal{X}_{\rm mercator}} \,,
\end{align}
where $\mathcal{X}\in \{ \mathcal{S}, s, R\}$ and the subscript indicates the embedding method.
Lower values of $\mathcal{S}$ and $R$ indicate higher-quality embeddings, while a higher value of $s$ indicates more accurate results.
The cases where dimensional annealing exhibits a better result are depicted in blue, and cases where Mercator shows a better result are depicted in red (Fig.~\ref{fig:synthetic_networks}).
In terms of NLL and greedy navigation success rate, dimensional annealing outperforms Mercator in all synthetic cases.
For the expansion ratio of the A* search algorithm, dimensional annealing achieves better results in 11 out of 15 cases.

In Fig.~\ref{fig:real_networks}, the same metrics are applied to real-world network datasets from social, biological, informational, and transportation domains.
Detailed information and the original source of the datasets are provided in Supplementary Table~S1.
The composition of panels and color schemes is the same as in Fig.~\ref{fig:synthetic_networks}.
In all three metrics, dimensional annealing outperforms Mercator in most cases.
Dimensional annealing yields a lower NLL than Mercator in 22 out of 25 cases.
The networks where Mercator exhibited better results are those with low $\beta \simeq 1$, where the networks exhibit weak or negligible latent geometric structure. 
Dimensional annealing achieves a higher greedy navigation success rate in 19 out of 25 networks.
Dimensional annealing achieves a more effective A* expansion ratio for 18 out of 25 networks.

\subsection{Scalability}

\begin{figure}[tb]
    \centering
    \includegraphics[width=0.85\columnwidth]{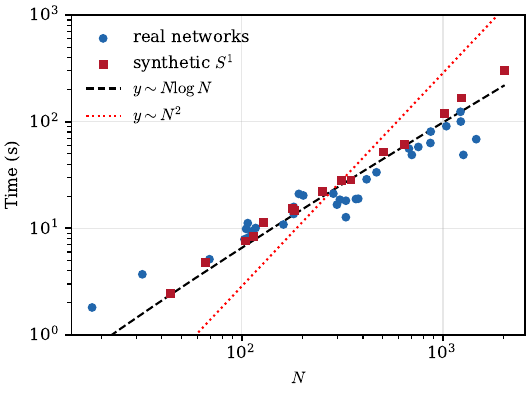}
    \caption{
    Embedding runtime as a function of network size $N$.
    Blue circles denote real networks, and the red squares denote synthetic $\mathbb{S}^1$ model networks.
    Dashed black line depicts $t \sim N \log N$, and dotted red line denotes $t \sim N^2$.
    Computational time is roughly proportional to $N \log N$.}
    \label{fig:runtime}
\end{figure}

Figure~\ref{fig:runtime} shows the runtime of dimensional annealing as a function of network size $N$.
Blue circles denote real networks and red squares depict synthetic $\mathbb{S}^1$ model networks.
For comparison, $y \propto N \log N$ is depicted as a dashed black line and $y\propto N^2$ as a dotted red line.
The data points approximately follow the dashed black line, and clearly do not scale as the dotted red line.
This contrasts with the $\mathcal{O}(N^2)$ computational scaling of the Mercator method~\cite{garcia2019mercator}.
Computational times were measured with an AMD Ryzen AI 9 HX 370 processor.

\section{Conclusion}
\label{sec:conclusion}

In this work, we introduced the \emph{dimensional annealing} framework combined with the Barnes--Hut approximation to obtain scalable, high-quality hyperbolic embeddings.
By introducing an auxiliary dimension in the latent space, topological constraints imposed by the one-dimensional similarity space can be bypassed.
The gradually increasing potential progressively confines the auxiliary coordinates, eventually recovering the original hyperbolic embedding.
This method is analogous to simulated annealing~\cite{kirkpatrick1983optimization} in many respects.
Simulated annealing gradually lowers the temperature while stochastically accepting uphill moves with temperature-dependent probability.
Dimensional annealing \emph{anneals the effective dimension}.
The potential barriers can be bypassed by exploiting the extra degree of freedom.
As optimization proceeds, the extra degree of freedom is progressively suppressed and eventually eliminated.
We showed that dimensional annealing produces high-quality embeddings across synthetic and real-world networks, as evaluated by NLL, greedy navigation success rate, and A* expansion ratio.
The Barnes--Hut approximation reduces the computational complexity from $\mathcal{O}(N^2)$ to $\mathcal{O}(N \log N)$.
The dimensional annealing framework may also be applicable to other inverse problems characterized by numerous local minima and topological constraints imposed by low dimensionality.
The present framework assumes a similarity space of $\mathbb{S}^1$; however, real networks may possess a latent space of higher dimension~\cite{budel2024random,almagro2022detecting,desy2023dimension} or nontrivial topology~\cite {jhun2022topological,taylor2015topological,mahler2021analysis}.
Extending the framework to such general latent spaces therefore provides a natural direction for future work.

\section{Acknowledgments}
We thank Deok-Sun Lee for many useful suggestions and discussions. This research was supported by the National Research Foundation of Korea (NRF) under grants RS-2026-25468383 (S.H.L.), RS-2023-00245174 (B.J.), RS-2025-00556024 (H.C.), and RS-2026-25539287 (G.S.); and by a KIAS Individual Grant (No. CG098002) at the Korea Institute for Advanced Study (G.S.).

\section{Code availability}

The hyperbolic embedding framework presented in this work is implemented in the \textbf{D}imensional \textbf{A}nnealing--based \textbf{Hy}perbolic \textbf{E}mbedding (DAHyE) library, written in C++ with a Python interface via \texttt{pybind11}.
The source code is publicly available at \href{https://github.com/bukyoungjhun/dahye}{https://github.com/bukyoungjhun/dahye}.

\bibliographystyle{elsarticle-num}
\bibliography{references}

\end{document}


\begin{frontmatter}

\title{Supplementary material: Hyperbolic network embedding via dimensional annealing and Barnes--Hut approximation}

\author[c3s,ceu,snu]{Bukyoung Jhun}
\ead{jhun@c3s.uni-frankfurt.de}
\author[KIAS,sap]{Gangmin Son}
\author[KIAS]{Hoyun Choi}
\author[GNU]{Sang Hoon Lee}

\address[c3s]{Center for Critical Computational Studies, Goethe-University Frankfurt, Frankfurt am Main, Germany}
\address[ceu]{Department of Network and Data Science, Central European University, Vienna, Austria}
\address[snu]{Department of Physics and Astronomy, Seoul National University, Seoul, Korea}
\address[KIAS]{School of Computational Science, Korea Institute for Advanced Study, Seoul, Korea}
\address[sap]{Dipartimento di Fisica, Sapienza Universit\`a di Roma, Rome, Italy}
\address[GNU]{Department of Physics, Gyeongsang National University, Jinju, Korea}
\ead{lshlj82@gnu.ac.kr}

\end{frontmatter}

\section{Dataset descriptions}
\label{sec:datasets}

Table~\ref{tab:network-descriptions} lists the empirical networks used in this study.
All networks were converted to undirected graphs with no self-loops or multiple edges before analysis.
For directed networks, edge directions were ignored, and for weighted networks, edge weights were ignored.
The \emph{Name} column gives the name of the network displayed in the main article, and the \emph{Key} column gives the identifier in the Netzschleuder database~\cite{peixoto2020netzschleuder}.
Rows follow the order in the figure, sorted by increasing
number of nodes. $N$ is the number of
nodes and $E$ is the number of edges. The analyzed
networks are undirected simple graphs.

\begingroup
\fontsize{9}{10}\selectfont
\setlength{\tabcolsep}{4pt}
\setlength{\LTcapwidth}{\textwidth}
\renewcommand{\arraystretch}{1.02}
\begin{longtable}{@{}>{\raggedright\arraybackslash}p{0.14\textwidth} >{\raggedright\arraybackslash}p{0.19\textwidth} >{\raggedright\arraybackslash}p{0.34\textwidth} r r >{\raggedright\arraybackslash}p{0.14\textwidth}@{}}
\caption{Empirical network data sets used in the real-network comparison.}\label{tab:network-descriptions}\\
\hline\hline
Name & Key & \multicolumn{1}{c}{Description} & $N$ & $E$ & Domain \\
\hline
\endfirsthead
\multicolumn{6}{c}{Table~\thetable\ (continued)}\\
\hline\hline
Name & Key & \multicolumn{1}{c}{Description} & $N$ & $E$ & Domain \\
\hline
\endhead
\hline
\endfoot
\hline
\endlastfoot
Malaria genes A & \texttt{malaria\_\allowbreak{}genes/\allowbreak{}HVR\_\allowbreak{}3} & Networks of recombinant antigen genes from the human malaria parasite \textit{P. falciparum}~\cite{Larremore_2013}. & 18 & 55 & Biological \\[2pt]
Highschool A & \texttt{add\_\allowbreak{}health/\allowbreak{}comm3} & Friendship network of U.S. high school students from the 1994–95 ADD Health survey.~\cite{Moody_2001}. & 32 & 91 & Social \\[2pt]
Highschool B & \texttt{add\_\allowbreak{}health/\allowbreak{}comm1} & Friendship network of U.S. high school students from the 1994–95 ADD Health survey~\cite{Moody_2001}. & 69 & 220 & Social \\[2pt]
Highschool C & \texttt{add\_\allowbreak{}health/\allowbreak{}comm2} & Friendship network of U.S. high school students from the 1994–95 ADD Health survey~\cite{Moody_2001}. & 103 & 348 & Social \\[2pt]
Political books & \texttt{polbooks} &Network of U.S. political books sold on Amazon around the 2004 presidential election, with edges representing frequent copurchases~\cite{krebs2004political}. & 105 & 441 & Informational \\[2pt]
Game of Thrones & \texttt{game\_\allowbreak{}thrones} & A coappearance network of Game of Thrones characters in A Storm of Swords, with edge weights indicating how often two characters’ names occur within 15 words of each other~\cite{beveridge2016network}. & 107 & 352 & Social \\[2pt]
Malaria genes B & \texttt{malaria\_\allowbreak{}genes/\allowbreak{}HVR\_\allowbreak{}2} & Networks of recombinant antigen genes from the human malaria parasite \textit{P. falciparum}~\cite{Larremore_2013}. & 112 & 743 & Biological \\[2pt]
Physicians & \texttt{physician\_\allowbreak{}trust} & Network of trust relationships among physicians in four Midwestern (USA) cities in 1966~\cite{Coleman_1957}. & 117 & 465 & Social \\[2pt]
Marvel & \texttt{marvel\_\allowbreak{}partnerships} & Network of partnerships among Marvel characters, extracted from Wikipedia, with nodes representing heroes or villains and edges representing partnerships. & 181 & 224 & Social\\[2pt]
C. elegans neural & \texttt{celegansneural} & Network representing the neural connections of the Caenorhabditis elegans nematode~\cite{white1986structure}. & 297 & 2148 & Biological \\[2pt]
Malaria genes C & \texttt{malaria\_\allowbreak{}genes/\allowbreak{}HVR\_\allowbreak{}1} & Networks of recombinant antigen genes from the human malaria parasite \textit{P. falciparum}~\cite{Larremore_2013}. & 307 & 2812 & Biological \\[2pt]
Facebook & \texttt{facebook\_\allowbreak{}friends} & A small anonymized Facebook ego network, from April 2014. Nodes are Facebook profiles, and an edge exists if the two profiles are ``friends'' on Facebook~\cite{maier2017cover}. & 329 & 1954 & Social \\[2pt]
\textit{E. coli} operons & \texttt{ecoli\_\allowbreak{}transcription/\allowbreak{}v1.0} & Network of operons and their pairwise interactions for \textit{E. coli}~\cite{Shen_Orr_2002}. & 329 & 456 & Biological \\[2pt]
London transport & \texttt{london\_\allowbreak{}transport} & Multiplex network with three edge types representing links within the three layers of London train stations: Underground, Overground, and DLR~\cite{2014}. & 369 & 430 & Transportation \\[2pt]
Netscience & \texttt{netscience} & A coauthorship network among scientists working on network science, from 2006. This network is a one-mode projection from the bipartite graph of authors and their scientific publications~\cite{Newman_2006}. & 379 & 914 & Social \\[2pt]
Euro airlines & \texttt{eu\_\allowbreak{}airlines} & A multiplex network of airline routes among European airports, where each of the 37 edge types represents routes by a different airline~\cite{2013}. & 417 & 2953 & Transportation \\[2pt]
Vienna streets & \texttt{urban\_\allowbreak{}streets/\allowbreak{}vienna} & Urban street networks, corresponding to 1-square-mile maps of 20 cities around the world~\cite{2006,2017}. & 467 & 691 & Transportation \\[2pt]
Wikipedia (Science) & \texttt{wiki\_\allowbreak{}science} & Network of scientific fields, extracted from the English Wikipedia in early 2020~\cite{Calderone2020}. & 677 & 6517 & Informational \\[2pt]
Internet & \texttt{internet\_\allowbreak{}top\_\allowbreak{}pop/\allowbreak{}Kdl} & Assorted snapshots of internet graph at the point of presence (PoP) level (which lies between the IP and AS levels), collected from around the world and at various times. The earliest snapshots are for ARPANET (1969--1972), with a few more from before 2000~\cite{Knight_2011}. & 754 & 895 & Technological \\[2pt]
Export products & \texttt{product\_\allowbreak{}space/\allowbreak{}HS} & Product network based on global trade, where edges connect products exported by similar countries and are weighted by export similarity~\cite{hidalgo2009building}. & 866 & 2532 & Economic \\[2pt]
Seoul streets & \texttt{urban\_\allowbreak{}streets/\allowbreak{}seoul} & Urban street networks, corresponding to 1-square-mile maps of 20 cities around the world~\cite{2006,2017}. & 869 & 1307 & Transportation \\[2pt]
Euro. roads & \texttt{euroroad} & Network of international ``E-roads,'' mostly in Europe~\cite{_ubelj_2011}. & 1039 & 1305 & Transportation \\[2pt]
Political blogs & \texttt{polblogs} & A directed network of hyperlinks among US political blogs~\cite{Adamic_2005}. & 1222 & 16714 & Informational \\[2pt]
Flight routes & \texttt{faa\_\allowbreak{}routes} & Network of air traffic routes, from the U.S. FAA (Federal Aviation Administration) National Flight Data Center (NFDC) preferred routes database~\cite{faa_routes}. & 1226 & 2408 & Transportation \\[2pt]
Scientific collab. & \texttt{new\_\allowbreak{}zealand\_\allowbreak{}collab} & Network of scientific collaborations among institutions in New Zealand~\cite{Aref_2018}. & 1463 & 4246 & Social \\[2pt]
\end{longtable}
\endgroup
